\documentclass{PoS}

\title{Two-pion scattering amplitude from Bethe-Salpeter wave function at the interaction boundary}

\ShortTitle{Two-pion scattering amplitude from BS wave function
at the interaction boundary}

\author{\speaker{Takeshi Yamazaki}$^{a,b}$\thanks{E-mail: yamazaki@het.ph.tsukuba.ac.jp}\hspace{2mm}  and Yusuke Namekawa$^{c}$\vspace{2mm}
\\
\llap{$^a$}
Faculty of Pure and Applied Sciences,
University of Tsukuba, Tsukuba, Ibaraki 305-8571, Japan\\
\llap{$^b$}
Center for Computational Sciences, University of Tsukuba,
Tsukuba, Ibaraki 305-8577, Japan\\
\llap{$^c$}
Institute of Particle and Nuclear Studies, High Energy Accelerator Research Organization(KEK), Tsukuba, Ibaraki 305-0801, Japan
}

\abstract{We observe that the ratio of the on-shell scattering amplitude to 
the Bethe-Salpeter (BS) wave function outside the interaction range 
is almost independent of time in our quenched calculation of the $I=2$ 
two-pion scattering with almost zero momentum. In order to discuss 
the time independence, we present a relation between the two-pion 
scattering amplitude and the surface term of the BS wave function 
at the boundary. Using the relation under some assumptions, 
we show that the ratio is independent of time if
the two-pion four-point function in early time is dominated by 
scattering states with almost zero momentum in addition to 
the ground state of the two-pion scattering.}

\FullConference{37th International Symposium on Lattice Field Theory - Lattice2019\\
		16-22 June 2019\\
		Wuhan, China}

\begin{document}

\section{Introduction}

The finite volume method derived by L\"uscher~\cite{Luscher:1990ux}
is widely employed for calculations of the scattering phase shift
$\delta(k)$ in lattice QCD.
This method is based on a relation between $\delta(k)$ and
the two-particle energy on finite volume of $L^3$.
The relation is originally derived in quantum mechanics~\cite{Luscher:1990ux},
and then the same relation is obtained in
quantum field theory using the BS wave function~\cite{Lin:2001ek,Aoki:2005uf}.
While in the derivation of the relation the BS wave function outside 
the interaction range $R$ is discussed,
the one inside $R$ in the infinite volume is also related to $\delta(k)$
through the on-shell scattering amplitude~\cite{Aoki:2005uf}.
The half-off-shell scattering amplitude can be defined in 
a similar way~\cite{Yamazaki:2017gjl}.

We extend the relation of the BS wave function inside $R$
in the infinite volume to the one on finite volume,
and perform an exploratory study using the extended relation in
the $I=2$ two-pion scattering with a small relative on-shell momentum $k$
at a heavy pion mass $m_\pi$ in
the quenched QCD~\cite{Namekawa:2017sxs}.
It is found that the results of $\delta(k)$ obtained from the two methods,
the finite volume method and the extended relation,
completely agree with each other,
and the half-off-shell amplitude can be calculated in a wide range of the momentum
with reasonable statistical error.
Furthermore, we confirm that similar results for the on-shell and half-off-shell
amplitudes are obtained in smaller $m_\pi$~\cite{Namekawa:2019xiy}.

\begin{figure}[h]
\center
\includegraphics*[scale=0.40]{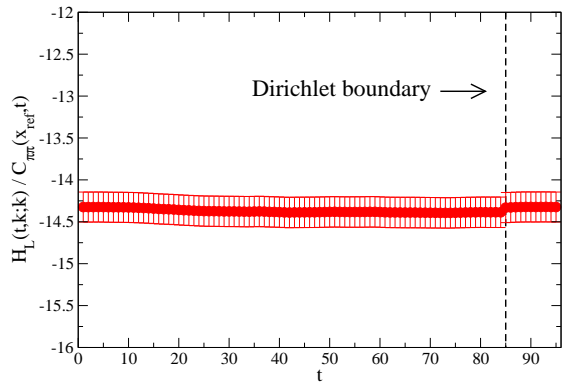}
\caption{
Ratio of time dependent on-shell amplitude on the lattice $H_L(t,k;k)$ to 
the two-pion four-point function $C_{\pi\pi}({\bf x}_{\rm ref},t)$ at
a reference point ${\bf x}_{\rm ref}$ ($x_{\rm ref} > R$) as a function of $t$.
The vertical dashed line represents the time slice of 
the Dirichlet boundary condition.
\label{fig:ratio}
}
\end{figure}

\begin{figure}[h]
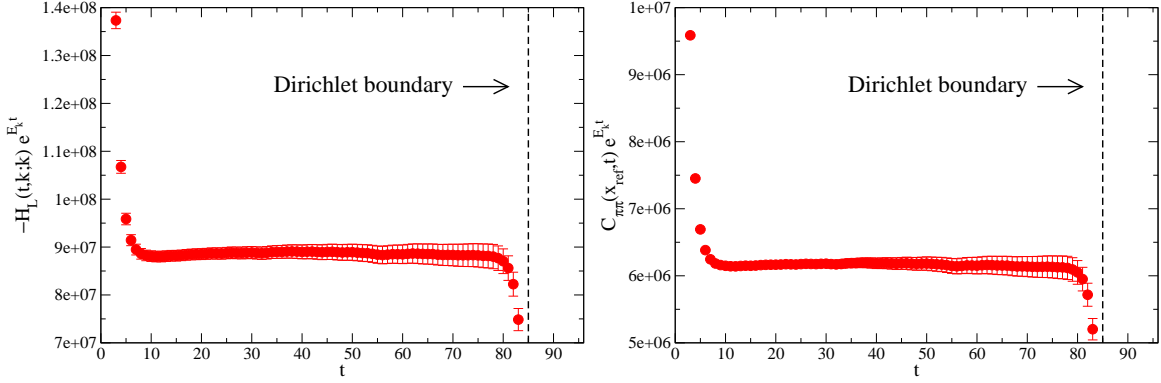

\includegraphics*[scale=0.40]{Figs/htkk.eps}
\includegraphics*[scale=0.40]{Figs/cxt.eps}
\caption{
Time dependences for $H_L(t,k;k)$ (Left) and $C_{\pi\pi}({\bf x}_{\rm ref},t)$ (Right)
normalized by $e^{E_k t}$ with the energy of the two-pion ground state 
$E_k$.
The vertical dashed line represents the time slice of 
the Dirichlet boundary condition.
\label{fig:ratio_each}
}
\end{figure}

In the study, we find that
a ratio of the time dependent on-shell amplitude on the lattice $H_L(t,k;k)$ 
to the four-point function $C_{\pi\pi}({\bf x}_{\rm ref},t)$ 
at a reference position ${\bf x}_{\rm ref}$ with 
$x_{\rm ref} = |{\bf x}_{\rm ref}| > R$
is independent of time $t$ as shown in Fig.~\ref{fig:ratio}.
This behavior is interesting, because
the numerator and denominator have significant $t$ dependences in small $t$
region. 
Figure~\ref{fig:ratio_each} presents the $t$ dependences for the numerator
and denominator normalized by the trivial exponential $t$ dependence
of the ground state.
In this report, we discuss conditions for 
this $t$ independence through
a definition of the scattering amplitude on the lattice
under some assumptions.
The results in this report have already been
presented in our paper~\cite{Namekawa:2019xiy}.

\section{Definitions}

In this report the S-wave two-pion scattering in
the center of mass frame is considered.
We assume that the interaction range is smaller than half of the 
spatial extent, $R < L/2$, and effects of inelastic scatterings are negligible.
The half-off-shell amplitude $H_L(p;k)$ is defined by 
the BS wave function of the two-pion ground state
on the lattice $\phi({\bf x};k)$~\cite{Namekawa:2017sxs},
\begin{equation}
 H_L(p;k) = -\sum_{{\bf x}} 
j_0(p x) (\Delta + k^2) \phi({\bf x};k) ,
\label{eq:def_off-shell}
\end{equation}
where $k^2 = (E_k^2 - 4m_\pi^2)/4$ with the energy 
of two-pion ground state $E_k$, and $\Delta$ is the symmetric Laplacian 
on the lattice. 
$j_0(p x)$ is the spherical Bessel function.
From a ratio of the on-shell amplitude $H_L(k;k)$ to
$\phi({\bf x}_{\rm ref};k)$ in $x_{\rm ref} > R$, 
$\delta(k)$ is obtained through the following relation,
\begin{equation}
\frac{H_L(k;k)}{\phi({\bf x}_{\rm ref};k)} =
\frac{4\pi x_{\rm ref} \sin\delta(k)}{\sin(k x_{\rm ref} + \delta(k))} .
\label{eq:on-shell_delta}
\end{equation}
It is assumed that contributions for higher angular 
momenta of $l \ge 4$ in $\phi({\bf x}_{\rm ref};k)$ are negligible
in the equation.

$H_L(p;k)$ can be written by a surface term using the partial integration
of Eq.~(\ref{eq:def_off-shell}), 
\begin{equation}
H_L(p;k) = -\sum_{{\bf x}} [(\Delta + k^2) j_0(p x)] \phi({\bf x};k) 
+ {\rm surf}(p;k)
\label{eq:off-shell_amp_surf}
\end{equation} 
with
\begin{equation}
{\rm surf}(p;k) = -3 \sum_{x_{1,2} = -L_{\rm min}}^{L_{\rm max}}
[ j_0(pX^\prime(L_{\rm min}))-j_0(pX^\prime(L_{\rm max}+1)) ]
\phi({\bf X}^\prime(L_{\rm max});k) ,
 \label{eq:surface_j0}
\end{equation}
where ${\bf X}^\prime(a) = (x_1,x_2,a)$, $L_{\rm max} = L/2$, and
$L_{\rm min} = L/2 - 1$.
Using this expression $H_L(k;k)$ is essentially given by
the surface term, in other words, $\phi({\bf x};k)$ at
the boundary on the lattice, because 
$(\Delta + k^2) j_0(k x) \approx 0$ in the small $k^2$ as in our calculation.
It is noted that since the summation in Eq.~(\ref{eq:def_off-shell}) 
can be replaced by the one inside $R$ 
because of $(\Delta + k^2) \phi({\bf x};k) = 0$
in $x > R$, $H_L(k;k)$ is expressed by $\phi({\bf x};k)$
at the interaction boundary with a different form 
of ${\rm surf}(k;k)$~\cite{Namekawa:2019xiy}.
A similar relation to Eq.~(\ref{eq:def_off-shell}) is obtained in the infinite
volume as discussed in Ref.~\cite{Carbonell:2016ekx}.

The time dependent amplitude on the lattice $H_L(t,k;k)$ 
is defined by the two-pion four-point function $C_{\pi\pi}({\bf x},t)$ as
\begin{equation}
 H_L(t,k;k) = -\sum_{{\bf x}} 
j_0(k x) (\Delta + k^2) C_{\pi\pi}({\bf x},t) ,
\end{equation}
where
\begin{equation}
C_{\pi\pi}({\bf x},t) = \langle 0 | \Phi({\bf x},t) \Omega^\dagger(0) | 0 \rangle .
\end{equation}
$\Omega^\dagger(0)$ is the two-pion source operator at $t=0$, where
each pion operator is projected to the zero momentum.
$\Phi({\bf x},t)$ is an $A_1^+$ projected two-pion operator,
\begin{equation}
\Phi({\bf x},t) = \sum_{\bf r} \pi^+(R_{A^+_1}[{\bf x}]+{\bf r},t)\pi^+({\bf r},t) ,
\end{equation}
where $R_{A^+_1}[{\bf x}]$ denotes the $A_1^+$ projection.

Since we discuss the $t$ dependences for $H_L(t,k;k)$ and $C_{\pi\pi}({\bf x},t)$
in small $t$ region, we consider $\pi\pi^\prime$ scattering states 
as well as the $\pi\pi$ scatterings in $C_{\pi\pi}({\bf x},t)$, 
where $\pi^\prime$ is the first radial excitation of $\pi$.
Then, $C_{\pi\pi}({\bf x},t)$ is written by those states as,
\begin{equation}
 C_{\pi\pi}({\bf x},t)
 = \sum_q A_q(t) \phi({\bf x};q)
 + \sum_{q^\prime} A^\prime_{q^\prime}(t)
     \phi({\bf x};q^\prime) 
= A_k(t) \phi({\bf x};k)( 1 + \delta C_{\pi\pi}({\bf x},t) ),
 \label{eq:C_pipi_exp}
\end{equation}
where $A_q(t) = C_q e^{-E_q t}$ and
$A^\prime_{q^\prime}(t) = C^\prime_{q^\prime}e^{-E^\prime_{q^\prime}t}$ 
with $E_q = 2 \sqrt{m_\pi^2 + q^2}$ and 
$E^\prime_q = \sqrt{m_\pi^2 + q^2} + \sqrt{m_{\pi^\prime}^2 + q^2}$. 
$C_{q}$ and $C^\prime_{q^\prime}$ are
overall constants, and
$\delta C_{\pi\pi}({\bf x},t)$ is all the excited state contributions 
divided by the ground state contribution of $\pi\pi$ scattering,
$A_k(t) \phi({\bf x};k)$.
In the $t \gg 1$ region, where the ground state dominates 
in $C_{\pi\pi}({\bf x},t)$,
the ratio $H_L(t,k;k)/C_{\pi\pi}({\bf x}_{\rm ref},t)$ is reduced to 
the ratio in Eq.~(\ref{eq:on-shell_delta}).
Note that it is straightforward to include $\pi^\prime\pi^\prime$ scattering states
in this discussion~\cite{Namekawa:2019xiy}.
Using a relation of ${\rm surf}(k;q)$,
\begin{equation}
- \sum_{\bf x} j_0(k x) (\Delta + k^2) \phi({\bf x};q)
= {\rm surf}(k;q) ,
\end{equation}
which is obtained using integration by parts 
under the assumption $(\Delta + k^2) j_0(kx) = 0$,
$H_L(t,k;k)$ is written in a similar form to Eq.~(\ref{eq:C_pipi_exp}),
$H_L(t,k;k) = A_k(t) H_L(k;k) ( 1 + \delta H_L(t,k;k) )$, where
$\delta H_L(t,k;k)$ is given by the sum of the surface terms for
the excited states,
\begin{equation}
\delta H_L(t,k;k) = 
\frac{\displaystyle{\sum_{q\ne k} A_q(t) {\rm surf}(k;q)
 + \sum_{q^\prime} A^\prime_{q^\prime}(t) {\rm surf}(k;q^\prime)}}
{A_k(t) H_L(k;k)} .
\label{eq:delta_H_L}
\end{equation}

\section{Time independence of $H_L(t,k;k) / C_{\pi\pi}({\bf x}_{\rm ref},t)$}

The $t$ dependence of the ratio 
$H_L(t,k;k) / C_{\pi\pi}({\bf x}_{\rm ref},t)$ is explained by
excited state contributions in Eqs.~(\ref{eq:C_pipi_exp}) 
and (\ref{eq:delta_H_L}),
\begin{equation}
 \frac{H_L(t,k;k)}{C_{\pi\pi}({\bf x}_{\rm ref},t)}
 =
 \frac{H_L(k;k)}{\phi({\bf x}_{\rm ref};k)}
 \frac{1 + \delta H_L(t,k;k)}{1 + \delta C_{\pi\pi}({\bf x}_{\rm ref},t)} .
 \label{eq:ratio_HL_phi}
\end{equation}
As shown in Fig.~\ref{fig:ratio}, this ratio behaves as a constant in $t$.
In the following we discuss sufficient conditions for the $t$ independence of the ratio 
using the surface term ${\rm surf}(k;q)$.

In a large $t$ region, the excited state parts decrease exponentially
compared to unity and the ground state dominates in both numerator
and denominator, so that the flat behavior in Fig.~\ref{fig:ratio}
is easy to understand.
On the other hand, in a small $t$ region,
it requires a nontrivial cancellation of the $t$ dependences 
in the excited state parts of Eq.~(\ref{eq:ratio_HL_phi}),
{\it i.e}, $\delta H_L(t,k;k) = \delta C_{\pi\pi}({\bf x}_{\rm ref},t)$.
In order to make the discussion simple,
we assume that the $t$ dependences reasonably coincide in each state,
in other words,
\begin{equation}
\frac{{\rm surf}(k;q)}{{\rm surf}(k;k)} \sim
\frac{\phi({\bf x}_{\rm ref};q)}{\phi({\bf x}_{\rm ref};k)} ,
\label{eq:condition}
\end{equation}
for each state with the momentum $q$, where we use $H_L(k;k) = {\rm surf}(k;k)$.

$\phi({\bf x};p)$ in $x > R$ is proportional to
the solution of the Helmholtz equation $G({\bf x};p)$ 
on finite volume~\cite{Luscher:1990ux},
\begin{equation}
G({\bf x};p) = \frac{1}{L^3}
     \sum_{{\bf q} \in \Gamma}  \frac{e^{i {\bf x} \cdot {\bf q} }}{q^2 - p^2},
 \label{eq:def_G}
\end{equation}
with 
\begin{equation}
\Gamma
 = \{ {\bf q} | {\bf q} = \frac{ 2 \pi }{ L } 
{\bf n}, {\bf n} \in {\bf Z}^3 \} .
\end{equation}
Furthermore, ${\rm surf}(k;p)$ can be evaluated
with $G({\bf x};p)$, because
${\rm surf}(k;p)$ is given by $\phi({\bf x};p)$
at the boundary, where $x > R$ is satisfied, as shown
in Eq.~(\ref{eq:surface_j0}).
Thus, the condition in Eq.~(\ref{eq:condition}) is rewritten 
in terms of $G({\bf x};p)$,
\begin{equation}
\frac{R_{\rm surf}(k;p)}{R_{\rm surf}(k;k)} \sim 1 ,
\end{equation}
where 
\begin{equation}
R_{\rm surf}(k;p) = \frac{{\rm surf}_G(k;p)}{G({\bf x}_{\rm ref};p)} .
\label{eq:def_R_surf}
\end{equation}
${\rm surf}_G(k;p)$ is the same as the surface term
in Eq.~(\ref{eq:surface_j0}),
while the BS wave function is replaced by $G({\bf X}^\prime(L_{\rm max});p)$.

\begin{figure}[t]
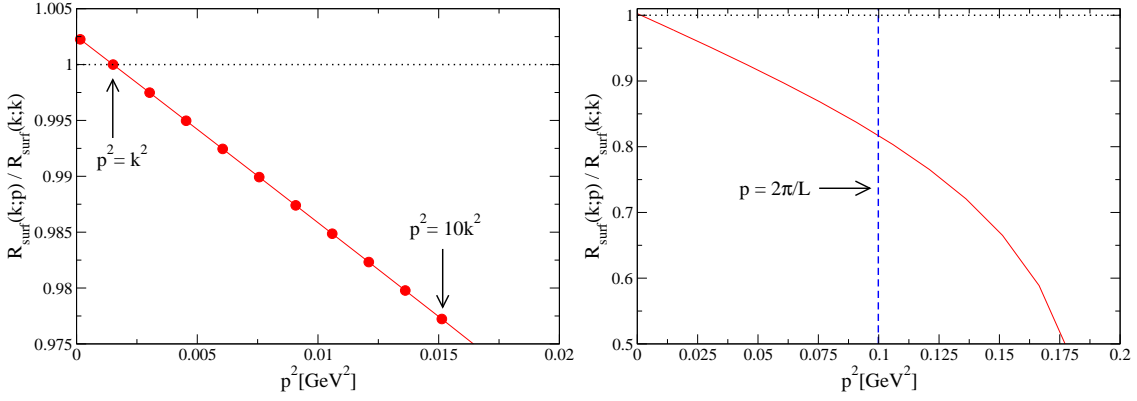

 \centering
 \includegraphics*[scale=0.40]{Figs/rat0.eps}
 \includegraphics*[scale=0.40]{Figs/rat1.eps}
 \caption{
  \label{fig:p2-ratio_surf_over_G}
  $p^2$ dependence of the ratio of
  $R_{\rm surf}(k;p) / R_{\rm surf}(k;k)$
  with $R_{\rm surf}(k;p)$ in Eq.~(\ref{eq:def_R_surf})
  at a reference position ${\bf x}_{\rm ref} = (12,7,2)$.
The left and right panels are the ratio in small and large momentum regions,
respectively.
The vertical dashed line in the right panel expresses $p = 2\pi/L$.
 }
\end{figure}

The ratio $R_{\rm surf}(k;p)/R_{\rm surf}(k;k)$ is plotted 
as a function of $p^2$ in Fig.~\ref{fig:p2-ratio_surf_over_G} with
a reference position ${\bf x}_{\rm ref} = (12,7,2)$.
The left panel presents that the ratio agrees with unity within
3\% in the small momentum region of $k^2 \le p^2 \le 10 k^2$.
This ratio decreases rapidly and significantly differs from unity 
near the non-zero smallest
momentum on finite volume, $p = 2\pi/L$,
denoted by the vertical dashed line in the right panel.

From the evaluation of $R_{\rm surf}(k;p)$
we conclude that the condition in Eq.~(\ref{eq:condition})
is reasonably satisfied for states with almost zero momentum,
whose value is similar to $k$.
In $C_{\pi\pi}({\bf x},t)$ such a state is identified as
the lowest $\pi\pi^\prime$ scattering state, which is expected to have
almost zero momentum.
Therefore, the sufficient condition for the $t$ independence of
$H_L(t,k;k) / C_{\pi\pi}({\bf x}_{\rm ref},t)$ is that
in small $t$ region
the lowest $\pi\pi^\prime$ scattering state has a large contribution
as well as the ground $\pi\pi$ state, 
and other excited states are negligible in
$H_L(t,k;k)$ and $C_{\pi\pi}({\bf x}_{\rm ref},t)$.
The conditions are reasonable, because
$C_{\pi\pi}({\bf x},t)$ is calculated using the zero momentum
projected $\pi$ operator, so that scattering states with
finite momentum are largely suppressed compared to the almost zero
momentum scattering states.
Furthermore, since we observe significant effect of the $\pi^\prime$ state
in the effective mass of the single $\pi$ correlator in $t < 10$,
it is expected that the lowest $\pi\pi^\prime$ state also has a large
contribution in $C_{\pi\pi}({\bf x},t)$ in the $t$ region.

\section{Summary}

We have observed an interesting behavior, that
the ratio $H_L(t,k;k) / C_{\pi\pi}({\bf x}_{\rm ref},t)$
is independent of $t$,
in the calculation of $\delta(k)$ from the BS wave function 
$\phi({\bf x};k)$ inside $R$.
To discuss sufficient conditions for the $t$ independence of the ratio,
we have derived another expression of the half-off-shell amplitude on the lattice,
which is written by the surface term, in other words,
the summation of $\phi({\bf x};k)$ at the boundary.
Not only $\pi\pi$ but also $\pi\pi^\prime$ scattering states
in $C_{\pi\pi}({\bf x},t)$ are considered, because 
we discuss $H_L(t,k;k) / C_{\pi\pi}({\bf x}_{\rm ref},t)$ in 
the small $t$ region.
Under an assumption that the $t$ dependences for 
$\delta H_L(t,k;k)$ and $\delta C_{\pi\pi}({\bf x}_{\rm ref},t)$
reasonably agree in each state,
we have obtained a condition given by the surface term and
the BS wave function.
The condition is examined by evaluating these quantities
using the solution of the Helmholtz equation on finite volume.
From the evaluation, we have found that 
the condition is satisfied when excited states have almost zero momentum.
Thus, the sufficient condition for the $t$ independence
of $H_L(t,k;k) / C_{\pi\pi}({\bf x}_{\rm ref},t)$ is that
$C_{\pi\pi}({\bf x},t)$ is dominated by the lowest $\pi\pi$
and $\pi\pi^\prime$ scattering states with almost zero momentum
in small $t$ region.
In order to confirm this condition 
analysis of $C_{\pi\pi}({\bf x},t)$ with more sophisticated method is
required, such as
the variational method~\cite{Luscher:1990ck}.

\section*{Acknowledgments}
We thank J.~Carbonell and V.~A.~Karmnov for pointing out 
the momentum space formulation of the on-shell amplitude,
and members of PACS collaboration for useful discussion.
Our simulation was performed on COMA
under Interdisciplinary Computational Science Program of Center for Computational Sciences, University of Tsukuba.
This work is based on Bridge++ code (http://bridge.kek.jp/Lattice-code/)~\cite{Ueda:2014rya}.
This work was supported in part by Grants-in-Aid 
for Scientific Research from the Ministry of Education, Culture, Sports, 
Science and Technology (Nos. 16H06002, 18K03638, 19H01892).

\bibliography{reference}

\end{document}